\documentclass[floats,prd,onecolumn,showpacs,amssymb]{revtex4}
\usepackage{amssymb}
\usepackage{graphics}
\usepackage{graphicx}
\usepackage{amsmath}
\usepackage{amsfonts}
\usepackage{bm}% bold math

\begin{document}

\title{ Constraining $f(T)$ teleparallel gravity by   Big Bang Nucleosynthesis}

\author{S. Capozziello$^{1,2,3}$, G. Lambiase$^{4,5}$, and E.N. Saridakis$^{6,7}$}%
%\email{}

\affiliation{$^1$Dipartimento di Fisica "E. Pancini", Universit\`{a} di Napoli
  ``Federico II'', Complesso Universitario di Monte Sant'Angelo,
  Edificio G, Via Cinthia, I-80126, Napoli, Italy
  \\
  $^2$Istituto Nazionale di Fisica Nucleare (INFN) Sezione
  di Napoli, Complesso Universitario di Monte Sant'Angelo, Edificio G,
  Via Cinthia, I-80126, Napoli, Italy
  \\
  $^3$ Gran Sasso Science Institute, Viale F. Crispi, 7, I-67100,
L'Aquila, Italy
\\
$^4$Dipartimento di Fisica E.R. Cainaiello, University of Salerno, Via
Giovanni Paolo
II, I 84084-Fisciano (SA), Italy\\
$^5$INFN, Gruppo Collegato di Salerno, Sezione di Napoli, Via Giovanni Paolo II, I
84084-Fisciano (SA), Italy\\
$^6$Department of
Physics, National Technical University of Athens, Zografou Campus
GR 157 73, Athens, Greece\\
$^7$CASPER, Physics Department, Baylor University, Waco, TX 76798-7310, USA}

%\date{August 2001}%
\def\be{\begin{equation}}
\def\ee{\end{equation}}
\def\al{\alpha}
\def\bea{\begin{eqnarray}}
\def\eea{\end{eqnarray}}

\begin{abstract}
We use BBN observational data on primordial abundance of ${}^4He$ to constrain
$f(T)$ gravity. The three most  studied viable $f(T)$ models, namely the
power law, the exponential and the
square-root exponential are considered, and  the BBN bounds are adopted in order to
extract
constraints on
their free parameters. For the power-law model, we find that the  constraints
are in agreement with those acquired using late-time cosmological data. For the
exponential and the square-root exponential models,  we show that for realiable
regions of  parameters space they always satisfy the BBN bounds. We conclude that
viable $f(T)$ models can successfully satisfy the BBN constraints.

\end{abstract}


\maketitle

\section{Introduction}

Cosmological observations coming from  Type Ia Supernovae \cite{Riess},  cosmic microwave
background
 radiation
\cite{Spergel} and the large scale structure \cite{Tegmark,Eisenstein}, provide
evidences that the Universe  is currently in an accelerating phase. This result is, in
general,
ascribed to the existence of a sort of  dark energy
(DE) sector in the universe, an exotic energy source characterized by a negative
pressure. At late times, the dark-energy sector eventually dominates over the cold dark
matter (CDM), and drives the Universe  to the observed accelerating expansion.

The simplest  candidate for DE is the cosmological constant $\Lambda$,
which has an equation-of-state parameter  $w=-1$. Although this model is in agreement with
current observations, it is plagued by some difficulties related to the small
observational value of DE density with respect to the expected one arising from quantum
field theories (the well known cosmological constant problem \cite{Carroll}). Moreover,
the $\Lambda$CDM paradigm, where cold dark matter (CDM) is considered into the game,  may
also
suffer from the age problem, as it was  shown
in \cite{Yang}, while the present data seem to slightly favor an evolving DE with the
 equation-of-state parameter crossing $w=-1$ from above to below in the near cosmological
past \cite{Feng}.

Over the past decade several DE models have been proposed, such as quintessence
\cite{Caldwell}, phantom \cite{Caldwell2}, k-essence \cite{Armendariz}, tachyon
\cite{Padmanabhan}, quintom \cite{Feng,Cai:2009zp,eli}, Chaplygin gas
\cite{Kamenshchik}, generalized Chaplygin gas (GCG) \cite{Bento}, holographic DE
\cite{Cohen,Li2004}, new agegraphic DE \cite{Wei},   Ricci DE \cite{Gao} etc. On the
other hand, there are also numerous models that induce an effective dark energy
which arises from modifications of the gravitational sector itself, such as
$f(R)$ gravity \cite{capfra,Nojiri,book,fRrev} (this class is very efficient in
verifying observational and theoretical constraints and explain the Universe acceleration
and phantom crossing \cite{noj1,noj2,noj4,noj5}), or gravity with higher curvature
invariants \cite{Mota}, by coupling the Ricci scalar to a scalar field
\cite{Farajollahi}, by introducing a vector field contribution \cite{Zuntz}, or by using
properties of gravity in higher dimensional spacetimes
\cite{Camera} (for a review see \cite{noj3,Capozziello:2011et}).

A possibility that can be explored to explain the accelerated phase of the Universe  is to
consider a theory of gravity based on the Weitzenb\"{o}ck connection, instead of the
Levi-Civita one, which deduces that the gravitational field is described by
the torsion instead of the curvature tensor. In such theories, the torsion tensor is
achieved from products of first derivatives of tetrad fields, and hence no second
derivatives appear. This  {\it Teleparallel}  approach \cite{Einstein,Einstein2},
is closely related to General Relativity, except for ``boundary terms'' \cite{seb1,seb2}
that involve
total
derivatives in the action, and thus one can construct the Teleparallel Equivalent of
General Relativity (TEGR), which is completely equivalent with General Relativity at the
level of equations but is based on torsion instead of  curvature. Teleparallel gravity
possesses a
number of attractive features
related to  geometrical and physical aspects \cite{Hayashi,Pereira.book,Maluf:2013gaa,
krssak}.
Hence, one can start from TEGR and construct various gravitational modifications based on
torsion, with $f(T)$ gravity being the most studied one
\cite{Ferraro2,Ferraro3,Linder:2010py}. In particular, it may
represent an alternative to inflationary models without the use of the inflaton, as well
as to effective DE models, in which the Universe acceleration is driven by the
extra torsion terms
\cite{Ferraro2,Ferraro3,Linder:2010py,Wu1,Chen:2010va,pertft,Dent:2011zz,Bamba:2010wb,Wu,
Zhang:2011qp,Bengochea,Yang2,Cai11,Li:2011rn,Bamba:2011pz,Daouda:2011rt,Atazadeh:2011aa,
Karami:2012fu,Cardone:2012xq,Otalora:2013tba,Ong:2013qja,Haro:2014wha,Harko:2014sja,
Hanafy:2014ica,Capozziello:2015rda,Bahamonde:2015zma,Carloni:2015lsa,Fazlpour:2016bxo}
(for a detailed review,  see \cite{Cai:2015emx}). The main advantage of
$f(T)$ gravity is that the field equations are 2nd-order ones, a property that
makes these theories simpler   if compared to the dynamical equations of other extended
theories of gravity, such as $f(R)$ gravity.

The aim of this paper is to explore the implications of $f(T)$ gravity to the formation
of light elements in the early Universe,  i.e. to the Big Bang Nucleosynthesis (BBN).
On the other hand, we want to explore the possibility to constrain $f(T)$  gravity by BBN
observatio
nal data.

BBN has occurred between the first fractions of  second after the Big Bang, around $\sim
0.01$ sec, and a few hundreds of seconds after it, when the Universe  was hot and
dense (indeed BBN, together with cosmic microwave background radiation, provides the
strong evidence about the high temperatures characterizing the primordial Universe ). It
describes the sequence of nuclear reactions that yielded the synthesis of light elements
\cite{kolb,bernstein}, and therefore drove the observed Universe. In general, from BBN
physics,  one may infer stringent constraints on a given cosmological model. Hence, in
this
work,  we shall confront various  $f(T)$ gravity models with BBN calculations based on
current observational data on primordial abundance of ${}^4He$, and
we shall  extract constraints on their free parameters.

The layout of the paper is as follows. In Section \ref{revmodel} we review $f(T)$ gravity
and the related cosmological models. In Section \ref{BBNanal} we use BBN
calculations in order to impose constraints on the free parameters of specific $f(T)$
gravity models. Conclusions are reported in Section \ref{Conclusions}. Finally,
in the Appendix we summarize the main notions of
BBN physics.

\section{$f(T)$ gravity and cosmology}
\label{revmodel}

Let us briefly review $f(T)$ gravity, and  apply it in a cosmological
framework. In this formulation,  the dynamical variable is the vierbein field
$e_i(x^\mu)$,
$i = 0, 1, 2, 3$, which forms an orthonormal basis in the tangent space at each point
$x^\mu$ of the manifold, i.e. $e_i \cdot e_j=\eta_{ij}$, with
$\eta_{ij}$ the Minkowsky metric with signature $-2$: $\eta_{ij}=diag(1,-1,-1,-1)$.
Denoting with $e^\mu_i$, $\mu=0,1,2,3$ the components of the vectors $e_i$ in a
coordinate
basis $\partial_\mu$, one can write $e_i=e^\mu_i\partial_\mu$. As a convection, here we
use the Latin indices for the tangent space, and the Greek indices for the coordinates on
the manifold. The dual vierbein allows to obtain the metric tensor of the manifold, namely
$g_{\mu\nu}(x)=\eta_{ij} e^i_\mu(x)e^j_\nu(x)$.

In teleparallel gravity, one adopts the curvatureless Weitzenb\"{o}ck connection
(contrarily to General Relativity which is based on the torsion-less Levi-Civita
connection), which gives rise to the non-null torsion tensor:
\begin{equation}\label{torsion}
T^\lambda_{\mu\nu}=\hat{\Gamma}^\lambda_{\nu\mu}-\hat{\Gamma}^\lambda_{\mu\nu}
=e^\lambda_i(\partial_\mu e^i_\nu - \partial_\nu e^i_\mu).
\end{equation}
Remarkably, the torsion tensor (\ref{torsion}) encompasses all the information about the
gravitational field. The Lagrangian density is built using its contractions,
%The Einstein field equations are a consequence of the
%dynamical
%equations for the vierbein.
and hence the teleparallel action is given by
 \begin{equation}\label{action0}
    I = \frac{1}{16\pi G}\int d^4x e T\,,
\end{equation}
with $e=det(e^i_\mu)=\sqrt{-g}$,
and where the torsion scalar $T$ reads as
\begin{equation}\label{lagrangian}
    T={S_\rho}^{\mu\nu}{T^\rho}_{\mu\nu}\,.
\end{equation}
Here, it is
\begin{eqnarray}
    {S_\rho}^{\mu\nu}&=&\frac{1}{2}({K^{\mu\nu}}_\rho+\delta^\mu_\rho
{T^{\theta\nu}}_\theta-\delta^\nu_\rho {T^{\theta\mu}}_\theta) \label{s} \\
    {K^{\mu\nu}}_\rho &=&
-\frac{1}{2}({T^{\mu\nu}}_\rho-{T^{\nu\mu}}_\rho-{T_\rho}^{\mu\nu})\,,
\label{contorsion}
\end{eqnarray}
with ${K^{\mu\nu}}_\rho$ the contorsion tensor which gives the difference between
Weitzenb\"{o}ck and Levi-Civita connections. Finally, the variation of action
(\ref{action0})
in terms of the vierbiens gives rise to the field equations, which coincide with those of
General Relativity. That is why the above theory is called the Teleparallel Equivalent of
General Relativity (TEGR).

One can now start from TEGR, and generalize action
(\ref{action0}) in order to construct gravitational modifications based on torsion. The
simplest scenario is to consider a Lagrangian density that is a function of
$T$, namely
\begin{equation}\label{action}
    I = \frac{1}{16\pi G}\int{d^4xe\left[T+f(T)\right]},
\end{equation}
that reduces to TEGR as soon as  $f(T)=0$. Considering additionally a matter Lagrangian
$L_m$, variation with respect to the vierbein gives the field equations
\cite{Cai:2015emx}
\begin{eqnarray}
\label{equations}
 &&e^{-1}\partial_\mu(e e_i^\rho {S_\rho}^{\mu\nu})[1+f']-
   e_i^\lambda {T^\rho}_{\mu\lambda}{S_\rho}^{\nu\mu}[1+f']\nonumber\\
   &&
  +e^\rho_i {S_\rho}^{\,\,\mu\nu}(\partial_\mu T)f'' +
    \frac{1}{4}e^\nu_i [T+f]=4\pi G\,{e_i}^\rho\, {\Theta_\rho}^\nu\,,
\end{eqnarray}
where $f'\equiv df/dT$, ${S_i}^{\mu\nu}={e_i}^\rho {S_\rho}^{\mu\nu}$ and
$\Theta_{\mu\nu}$ is the energy-momentum tensor for the matter sector.

In order to explore the cosmological implications of $f(T)$ gravity, we focus on
homogeneous and isotropic geometry, considering the usual choice for the vierbiens, namely
\begin{equation}
\label{weproudlyuse}
e_{\mu}^A={\rm diag}(1,a,a,a),
\end{equation}
which corresponds to a flat Friedmann-Robertson-Walker (FRW) background
metric of the form
\begin{equation}
ds^2= dt^2-a^2(t)\,\delta_{ij} dx^i dx^j,
\end{equation}
where $a(t)$ is the scale factor.
Equations (\ref{torsion}), (\ref{lagrangian}), (\ref{s}) and (\ref{contorsion}) allow to
derive a relation
between the torsion $T$ and
the Hubble parameter ${\displaystyle H=\frac{\dot a}{a}}$, namely
\begin{equation}\label{lt}
    T=-6H^2.
\end{equation}
Hence, in the case of FRW geometry, and assuming that the matter sector corresponds to a
perfect fluid with energy density $\rho$ and pressure $p$, the $i=0=\nu$ component of
(\ref{equations}) yields
\begin{equation}\label{friedmann}
    12H^2[1+f']+[T+f]=16\pi G\rho,
\end{equation}
while the $i=1=\nu$ component gives
\begin{equation}\label{acceleration}
    48H^2f''\dot{H}-(1+f')[12H^2+4\dot{H}]-(T-f)=16\pi Gp.
\end{equation}
The equations close by considering the equation of continuity for the matter sector,
namely
$\dot{\rho}+3H(\rho+p)=0$.
One can rewrite (\ref{friedmann}) and (\ref{acceleration}) in the usual form
\begin{equation}\label{modfri}
    H^2=\frac{8\pi G}{3}(\rho+\rho_T),
\end{equation}
\begin{equation}\label{modacce}
    2\dot H+3H^2=-\frac{8\pi G}{3}(p+p_T)\,,
\end{equation}
%from which one gets
% \begin{equation}\label{Hdot}
%   {\dot H}=-4\pi G \frac{\rho+p}{1+f'+2T f''}\,.
% \end{equation}
where
\begin{eqnarray}\label{rhoT}
    \rho_T &=& \frac{3}{8\pi G}\left[\frac{Tf'}{3}-\frac{f}{6}\right],\\
    p_T&=&\frac{1}{16\pi G}\, \frac{f-T f'+ 2T^2f''}{1+f'+ 2{T}f''} \,,
\label{pT}
\end{eqnarray}
are the effective energy density and pressure arising from torsional contributions. One
can therefore define the effective torsional  equation-of-state parameter  as
\begin{equation}\label{omegaeff}
    \omega_{T}\equiv \frac{p_T}{\rho_T}= -\frac{f-T f'+2{T}^2f''}{(1+f'+
2 T
f'')(f-2Tf')}\,.
\end{equation}
In these classes of theories, the above effective torsional terms are responsible for the
accelerated phases of the early or/and late Universe \cite{Cai:2015emx}.

Let us present now  three  specific $f(T)$ forms, which are the
viable ones amongst the variety of $f(T)$ models  with two parameters out of which
one is independent, i.e which pass the basic observational tests
\cite{Nesseris:2013jea}.

\begin{enumerate}

\item The power-law model by Bengochea and Ferraro
(hereafter $f_{1}$CDM) \cite{Ferraro3} is characterized by the form
    \begin{equation}\label{eq:ftmyrzabis}
f(T) = \beta |T|^n,
\end{equation}
where $\beta$ and $n$ are the two model parameters.
Inserting this $f(T)$ form into  Friedmann equation (\ref{friedmann}) at present,
we
acquire
\begin{eqnarray}
\beta=(6H_0^2)^{1-n}\frac{\Omega_{m0}}{2n-1},
\end{eqnarray}
 where   ${\displaystyle \Omega_{m0}=\frac{8\pi G \rho_{m}}{3H_0^2}}$ is the matter
density parameter at present, and
 \begin{eqnarray*}
H_0 &=& 73.02\pm 1.79\mbox{km/(sec Mpc)} \\
   &\sim & 2.1 \times 10^{-42}\mbox{GeV}
 \end{eqnarray*}
is the current Hubble parameter value. The best fit on the parameter $n$ is obtained
taking the $CC+H_0+SNeIa+BAO$ observational data, and it reads \cite{TJCAP16}
   \begin{equation}\label{bestfit1}
    n=0.05536\,.
  \end{equation}
Clearly, for $n=0$  the present scenario reduces to $\Lambda$CDM cosmology, namely
${T}+f(T)=T-2\Lambda$, with $\Lambda=-\beta/2$.

\item The Linder model (hereafter $f_{2}$CDM) \cite{Linder:2010py} arises from
\begin{eqnarray}
\label{modf2}
f(T)=\alpha T_{0}(1-e^{-p\sqrt{T/T_{0}}}),\quad
p=\frac{1}{b}\,,
\end{eqnarray}
with $\alpha$ and $p$ ($b$) the two model parameters. In this case
(\ref{friedmann})  gives that
\begin{eqnarray}
\alpha=\frac{\Omega_{m0}}{1-(1+p)e^{-p}}\,.
\end{eqnarray}
 The $CC+H_0+SNeIa+BAO$ observational data imply that the best fit of $b$ is
\cite{TJCAP16}
  \begin{equation}\label{bestfit2}
    b=0.04095\,.
  \end{equation}
As we can see,   for $p \rightarrow +\infty$ the present scenario reduces
to  $\Lambda$CDM cosmology.

 \item Motivated by  exponential $f(R)$ gravity \cite{Linder:2009jz}, Bamba et. al.
introduced the following   $f(T)$ model (hereafter $f_{3}$CDM)
\cite{Bamba:2010wb}:
 \begin{eqnarray}
 \label{modf3}
 f(T)=\alpha T_{0}(1-e^{-pT/T_{0}}), \quad p=\frac{1}{b}\,,
 \end{eqnarray}
 with $\alpha$ and $p$  ($b$) the two model parameters.
 In  this case we acquire
 \begin{eqnarray}
 \alpha=\frac{\Omega_{m0}}{1-(1+2p)e^{-p}}.
 \end{eqnarray}
 For this model, and using $CC+H_0+SNeIa+BAO$ observational data, the best fit is found
to be \cite{TJCAP16}
  \begin{equation}\label{bestfit3}
    b=0.03207\,.
  \end{equation}
  Similarly to the previous case  we can immediately see that    $f_{3}$CDM model tends
to   $\Lambda$CDM cosmology for $p \rightarrow +\infty$.
% or
%equivalently for $b
% \rightarrow 0^{+}$.

 \end{enumerate}

The above  $f(T)$ models are considered viable  in
literature because pass the basic observational tests \cite{Cai:2015emx}.  They are
characterized
by two free parameters. Actually there are two more models with two
free parameters, namely the logarithmic model  \cite{Bamba:2010wb},
\begin{equation}
f(T)=\alpha T_{0} \sqrt{\frac{T}{cT_{0}}}
\ln\left(\frac{cT_{0}}{T}\right )\,,
\end{equation}
  and the hyperbolic-tangent model
\cite{Wu},
\begin{equation}
f(T)=\alpha(-T)^{n}\tanh\left(\frac{T_{0}}{T}\right)\,.
\end{equation}
Nevertheless since these two models do not possess  $\Lambda$CDM
cosmology as a limiting case and since they are in tension with
observational data \cite{Nesseris:2013jea}, in this work
we do not consider them.

Finally, let us note that one could also
construct $f(T)$ models with more than two parameters, for example,  combining the
above  scenarios.  However, considering  many free parameters would be a
significant disadvantage concerning the corresponding values of the
information criteria.

\section{Big Bang Nucleosynthesis  in $f(T)$ cosmology}
\label{BBNanal}

In the Section,  we examine the BBN in the framework of
$f(T)$ cosmology. As it is well known, BBN occurs during the radiation
dominated era. The energy density of relativistic particles filling up the Universe  is
given by ${\displaystyle \rho=\frac{\pi^2}{30}g_* {\cal T}^4}$,
where $g_*\sim 10$ is
the effective number of degrees of freedom and ${\cal T}$ the temperature (in the
Appendix we review the main
features related to the BBN physics). The neutron abundance is computed via the
conversion rate of protons into neutrons, namely
 \[
 \lambda_{pn}({\cal T})=\lambda_{n+\nu_e\to p+e^-}+\lambda_{n+e^+\to p+{\bar
\nu}_e}+\lambda_{n\to p+e^- +
{\bar \nu}_e}\,,
\]
and its inverse $\lambda_{np}({\cal T})$. The relevant quantity is the total rate given by
 \begin{equation}\label{Lambda}
    \Lambda({\cal T})=\lambda_{np}({\cal T})+\lambda_{pn}({\cal T})\,.
 \end{equation}
Explicit calculations of Eq. (\ref{Lambda}) lead to (see (\ref{LambdafinApp}) in the
Appendix)
 \begin{equation}\label{Lambdafin}
    \Lambda({\cal T}) =4 A\, {\cal T}^3(4! {\cal T}^2+2\times 3! {\cal Q}{\cal T}+2!
{\cal Q}^2)\,,
 \end{equation}
where ${\cal Q}=m_n-m_p$ is the mass difference of neutron and proton, and $A=1.02 \times
10^{-11}$GeV$^{-4}$. The primordial mass
fraction of ${}
^4 He$ can be estimated by making use of the relation \cite{kolb}
 \begin{equation}\label{Yp}
    Y_p\equiv \lambda \, \frac{2 x(t_f)}{1+x(t_f)}\,.
 \end{equation}
Here $\lambda=e^{-(t_n-t_f)/\tau}$, with $t_f$   the time of the freeze-out of the weak
interactions, $t_n$   the time of the freeze-out of the nucleosynthesis,
$\tau$ the neutron mean lifetime given in (\ref{rateproc3}), and
$x(t_f)=e^{-{\cal
Q}/{\cal T}(t_f)}$ is the neutron-to-proton equilibrium ratio.
The function $\lambda(t_f)$ is interpreted as the fraction of neutrons that decay into
protons during the interval $t\in [t_f, t_n]$. Deviations from the fractional mass $Y_p$
due to
the variation of the freezing temperature ${\cal T}_f$ are given by
 \begin{equation}\label{deltaYp}
    \delta
Y_p=Y_p\left[\left(1-\frac{Y_p}{2\lambda}\right)\ln\left(\frac{2\lambda}{Y_p}
-1\right)-\frac{2t_f}{\tau}\right]
    \frac{\delta {\cal T}_f}{{\cal T}_f}\,,
 \end{equation}
where we have set $\delta {\cal T}(t_n)=0$ since ${\cal T}_n$ is fixed by the deuterium
binding energy
\cite{torres,Lambiase1,Lambiase2,Lambiase3}. The experimental estimations of the mass
fraction $Y_p$ of baryon
converted to ${}^4 He$
during the Big Bang Nucleosynthesis are
\cite{coc,altriBBN1,altriBBN2,altriBBN3,altriBBN4,altriBBN5,altriBBN6}
 \begin{equation}\label{Ypvalues}
 Y_p=0.2476\,, \qquad |\delta Y_p| < 10^{-4}\,.
 \end{equation}
 Inserting these into (\ref{deltaYp}) one infers the upper bound on $\frac{\delta
{\cal T}_f}{{\cal T}_f}$, namely
 \begin{equation}
 \label{deltaT/Tbound}
    \left|\frac{\delta {\cal T}_f}{{\cal T}_f}\right| < 4.7 \times 10^{-4}\,.
 \end{equation}

During the BBN, at the radiation dominated era, the scale factor evolves as $a\sim
t^{1/2}$, where $t$ is cosmic time. The  torsional energy density $\rho_T$ is
treated as a perturbation to the radiation energy density $\rho$. The relation between
the
cosmic time and the temperature
is given by ${\displaystyle \frac{1}{t}\simeq (\frac{32\pi^3 g_*}{90})^{1/2}\frac{{\cal
T}^2}{M_{P}}
}$
(or
${\cal T}(t)\simeq (t/\text{sec})^{1/2} $MeV). Furthermore, we use the entropy
conservation
$S\sim a^3
{\cal T}^3=constant$. The expansion rate of
the Universe  is derived from (\ref{modfri}), and can be rewritten in the form
  \begin{eqnarray}\label{H+H1}
    H&=&H_{GR}^{(R)}\sqrt{1+\frac{\rho_T}{\rho}}=H_{GR}+\delta H\,,  \\
    \delta H&=&\left(\sqrt{1+\frac{\rho_T}{\rho}}-1\right)H_{GR}\,,
 \end{eqnarray}
where $H_{GR}=\displaystyle{\sqrt{\frac{8\pi G}{2}\rho}}$ ($H_{GR}$ is the expansion rate
of the
Universe  in General Relativity). Thus, from the relation $\Lambda= H$, one derives the
freeze-out
temperature
${\cal T}={\cal T}_f\left(1+\frac{\delta {\cal T}_f}{{\cal T}_f}\right)$, with ${\cal
T}_f\sim 0.6$ MeV (which follows from
$H_{GR}\simeq q {\cal T}^5$)
and
\begin{equation}\label{H_T=Lambda}
  \left(\sqrt{1+\frac{\rho_T}{\rho}}-1\right)H_{GR} = 5q {\cal T}_f^4 \delta {\cal T}_f\,,
  \end{equation}
from which, in the regime $\rho_T\ll \rho$, one obtains:
  \begin{equation}\label{deltaT/TboundG}
  \frac{\delta {\cal T}_f}{{\cal T}_f}\simeq \frac{\rho_T}{\rho}\frac{H_{GR}}{10 q
{\cal T}_f^5}\,,
\end{equation}
with $q=4! A\simeq 9.6\times 10^{-36}$GeV$^{-4}$.

In what follows we shall investigate the bounds that arise from the BBN constraints, on
the free parameters of the three $f(T)$ models presented in the previous Section.
These constraint will be determined using Eqs. (\ref{deltaT/TboundG}) and (\ref{rhoT}).
Moreover, we shall use the   numerical values
 \[
 \Omega_{m0}=0.25\,, \quad {\cal T}_0=2.6\times 10^{-13}\mbox{GeV}\,,
 \]
where ${\cal T}_0$ is the present value of CMB temperature.

\begin{enumerate}

\item  $f_1$CDM model.

For the $f_1$CDM model of (\ref{eq:ftmyrzabis}) relation (\ref{rhoT}) gives
 \begin{eqnarray}
   \rho_T &=& \frac{1}{16\pi G}\left[\beta (2n-1)(|6H^2|)^n\right] \nonumber \\
   &=& \frac{3H_0^2}{8\pi G}\, \Omega_{m0}\left(\frac{{\cal T}}{{\cal T}_0}\right)^{4n}
\label{rhoTH}
\,,
 \end{eqnarray}
and then (\ref{deltaT/TboundG}) yields
%\begin{widetext}
 \begin{equation}\label{deltaT/TboundGFin}
   \frac{\delta {\cal T}_f}{{\cal T}_f}=\frac{\pi}{15}\sqrt{\frac{\pi
g_*}{5}}\,\Omega_{m0}\left(\frac{{\cal T}_f}{{\cal
T}_0}\right)^{4(n-1)}\frac{1}{qM_{Pl}{\cal T}_f^3}\,.
 \end{equation}
%\end{widetext}
In Fig. \ref{deltaTf1} we depict $\delta {\cal T}_f/{\cal T}_f$ from
(\ref{deltaT/TboundGFin})
vs $n$, as well as the upper bound from (\ref{deltaT/Tbound}). As we can see,
constraints from BBN require $n\lesssim 0.94$. Remarkably, this bound is in agreement with
  the best fit for $n$ of (\ref{bestfit1}), namely $n=0.05536$, that was obtained
using $CC+H_0+SNeIa+BAO$ observational data in \cite{TJCAP16}.
\begin{figure}[ht]
\includegraphics[width=2.9in,height=2.25in]{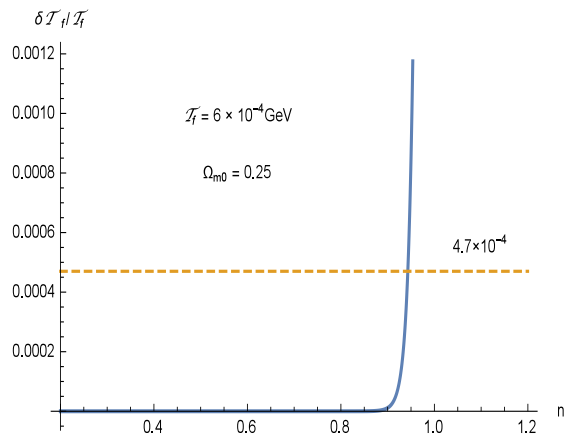}
\caption{\label{deltaTf1} \textit{ $\delta {\cal T}_f/{\cal T}_f$ from
(\ref{deltaT/TboundGFin})
vs $n$ (thick line) for the $f_1$CDM model of (\ref{eq:ftmyrzabis}), and the upper bound
for $\delta {\cal T}_f/{\cal T}_f$ from (\ref{deltaT/Tbound}) (dashed line). As we can
see,
constraints from BBN require $n\lesssim 0.94$.}}
\end{figure}

\item $f_{2,3}$CDM model.

In the case of $f_2$CDM model of (\ref{modf2}) and $f_3$CDM  model of (\ref{modf3}), and
for the purpose of this analysis, we can unified their investigation  parameterizing them
  as
      \begin{equation}\label{f12insieme}
      f(T)= \alpha T_0 \left[1-e^{-p(T/T_0)^m}\right]\,,
      \end{equation}
      with
      \[
      \alpha = \frac{\Omega_{m0}}{1-(1+2mp)e^{-p}}\,,
      \]
where $ m=\frac{1}{2}$ for model  $f_2$CDM  and $ m=1$ for model  $f_3$CDM.
Inserting (\ref{f12insieme}) into (\ref{deltaT/TboundG}) we acquire
\begin{eqnarray}\label{delta12insieme}
&&
\!\!\!\!\!
     \frac{\delta {\cal T}_f}{{\cal T}_f}=\frac{2\pi \alpha}{15}\sqrt{\frac{\pi g_*}{5}}
         \left(\frac{{\cal T}_0}{{\cal T}_f}\right)^4\frac{1}{qM_P {\cal
T}_f^3}\nonumber\\
 &&
\ \ \ \ \ \cdot
     \left\{
     \left[mp\left(\frac{{\cal
T}_0}{{\cal T}_f}\right)^{4m}\!+\!\frac{1}{2}\right]e^{-p({\cal T}_f/{\cal T}_0)^{4m}}
    \! -\!\frac{1}{2}\right\}.
\end{eqnarray}
Hence, using this relation we can calculate the value of $|\delta {\cal T}_f/{\cal T}_f|$
for various values of $p=1/b$ that span the order of magnitude of the best fit values
(\ref{bestfit1}) and (\ref{bestfit2}) that were obtained using $CC+H_0+SNeIa+BAO$
observational data in \cite{TJCAP16}, and we present our results in Table \ref{Tab1}.
As we can see, in all cases the value of   $|\delta {\cal T}_f/{\cal T}_f|$ is well below
the BBN bound  (\ref{deltaT/Tbound}). Hence, BBN cannot impose constraints on the
parameter values of $f_2$CDM and $f_3$CDM models.
\end{enumerate}
   \begin{center}
\begin{table}[ht]
\begin{tabular}{ccccc}
  \hline\hline
  $m$ & \qquad \qquad & $p=1/b$ & \qquad \qquad  & $|\delta {\cal T}_f/{\cal T}_f|$  \\
\hline
  1/2 & & 1   & & $5.723 \times 10^{-38}$ \\
      & & $10$ & & $1.512 \times 10^{-38}$ \\
      & &$10^2$ & & $1.511 \times 10^{-38}$ \\ \hline
    1 & & 1     & & $1.4586 \times 10^{-37}$ \\
      & & 10 & & $1.5131 \times 10^{-38}$ \\
      & & $10^2$ & &  $1.5116 \times 10^{-38}$ \\
  \hline\hline
 \end{tabular}
\caption{$|\delta {\cal T}_f/{\cal T}_f|$ from (\ref{delta12insieme}) for different
values of $p=1/b$, for  $m=1/2$ ($f_2$CDM model) and $m=1$ ($f_3$CDM model).
%obtained from the best fit of $\Lambda$CDM, using the
%$CC+H_0+SNeIa+BAO$
%observational
%data.
}
\label{Tab1}
\end{table}
\end{center}

\section{Conclusions}
\label{Conclusions}

In this work we have investigated the implications of $f(T)$ gravity to the formation
of light elements in the early Universe, i.e. to the BBN.
In particular, we have examined the three most used and well studied viable $f(T)$
models, namely the power law, the exponential and the square-root exponential, and we
have confronted them with BBN calculations based on current observational data on
primordial abundance of ${}^4He$. Hence, we were able to extract constraints on
their free parameters.

Concerning the power-law $f(T)$ model, the obtained constraint on the exponent $n$,
is $n\lesssim 0.94$. Remarkably, this bound is in agreement with the constraints
obtained using $CC+H_0+SNeIa+BAO$ observational data  \cite{TJCAP16}. Concerning the
exponential and the square-root exponential, we showed that, for  realistic regions of
free parameters,  they always satisfy the BBN bounds. This means that, in these cases,
BBN cannot
impose strict
constraints on the values  of  free parameters.

In summary,  we showed that  viable $f(T)$ models, namely those that pass
the basic observational tests, can also satisfy the BBN constraints. This feature acts as
an additional advantage of $f(T)$ gravity, which might be a successful candidate for
 describing the  gravitational interaction. As discussed in \cite{Cai:2015emx}, this kind
of
constraints could contribute in the debate of fixing the most realistic picture that can
be based
on curvature or torsion.

 \begin{acknowledgments}
This article is based upon work from COST Action CA15117 ``Cosmology and Astrophysics
Network
for Theoretical Advances and Training Actions'' (CANTATA), supported by COST (European
Cooperation
in Science and Technology).
 \end{acknowledgments}

\appendix*

\section{Big Bang Nucleosynthesis}

In this Appendix we briefly review the main features of Big Bang Nucleosynthesis following
\cite{kolb,bernstein}.
%The energy density of relativistic particles ($T\gg m, \mu$, where $\mu$ is the chemical
%potential)
% filling up the early Universe  is given by $\rho=\frac{g_s}{(2\pi)^2}\int E n(E/T)
%d^3p=\frac{\pi^2}
%{30}g T^4$.
%%, where $g_s$ represents the degeneracy factors for particle species ($g_\gamma=2$,
%$g_{e}=4$, $g_\nu=2$), and $g=g_b+\frac{7}{8}g_f=\frac{43}{4}$ ($g_f=g_e+3g_\nu=10$) is
%the effective %number of
%degree of freedom (it is implicitly assumed that muon and tau neutrinos have a small
%mass compared
%to the effective temperature and that no other massless species are present).
In the early Universe, the primordial ${}^4He$ was formed at temperature ${\cal T}\sim
100$ MeV. The energy and number density were formed by relativistic leptons (electron,
positron and neutrinos) and photons. The rapid collisions maintain all these particles in
thermal equilibrium. Interactions of protons and neutrons were kept in thermal
equilibrium by means of their interactions with leptons
 \begin{eqnarray}\label{proc1}
    \nu_e+n & \longleftrightarrow & p+e^- \\
    e^++n & \longleftrightarrow & p + {\bar \nu}_e \label{proc2} \\
    n& \longleftrightarrow & p+e^- + {\bar \nu}_e\,. \label{proc3}
 \end{eqnarray}

The neutron abundance is estimated by computing the conversion rate of protons into
neutrons, i.e. $\lambda_{pn}({\cal T})$, and its inverse $\lambda_{np}({\cal T})$. Thus,
the weak interaction rates (at suitably high temperature)
are given by
 \begin{equation}\label{LambdaA}
    \Lambda({\cal T})=\lambda_{np}({\cal T})+\lambda_{pn}({\cal T})\,.
 \end{equation}
The rate $\lambda_{np}$ is the sum of the rates associated to the processes
(\ref{proc1})-(\ref{proc3}), namely
 \begin{equation}\label{sumprocess}
    \lambda_{np}=\lambda_{n+\nu_e\to p+e^-}+\lambda_{n+e^+\to p+{\bar
\nu}_e}+\lambda_{n\to p+e^- +
{\bar \nu}_e}\,.
 \end{equation}
 Finally, the rate $\lambda_{np}$ is related to the rate $\lambda_{pn}$ as
$\lambda_{np}({\cal T})=e^{-{\cal Q}/{\cal T}}\lambda_{pn}({\cal T})$, with ${\cal
Q}=m_n-m_p$ the mass difference of neutron and proton.

During the freeze-out stage, one can use the following approximations
\cite{bernstein}: (i) The
temperatures of particles are the same, i.e. ${\cal T}_\nu={\cal T}_e={\cal
T}_\gamma={\cal T}$. (ii) The temperature
${\cal T}$ is lower than the typical energies $E$ that contribute to the
integrals entering the definition of the rates (one can therefore replace the
Fermi-Dirac distribution with the Boltzmann
one, namely $n\simeq e^{-E/{\cal T}}$). (iii) The electron mass $m_e$ can be neglected
with
respect to the
electron and neutrino energies ($m_e\ll E_e, E_\nu$).

Having these in mind, the interaction rate
corresponding to the process (\ref{proc1}) is given by
  \begin{equation}\label{rateproc1}
    d\lambda_{n+\nu_e\to p+e^-}= d\mu \,(2\pi)^4 |\langle{\cal M}|^2\rangle W \,,
  \end{equation}
where
\begin{eqnarray}
  d\mu & \equiv &  \frac{d^3p_e}{(2\pi)^3 2E_e} \frac{d^3p_{\nu_e}}{(2\pi)^3
2E_{\nu_e}}\frac{d^3p_
p}{(2\pi)^3 2E_p}\,, \label{dmu} \\
  W &\equiv & \delta^{(4)}({\cal P})n(E_{\nu_e})[1-n(E_e)]\,, \label{WA}\\
   {\cal P}  &   \equiv &  p_n+p_{\nu_e}-p_p-p_e\,,  \\
 {\cal M} &= &\left(\frac{g_w}{8M_W}\right)^2 [{\bar u}_p\Omega^\mu u_n][{\bar
u}_e\Sigma_\mu v_{\nu_e}]\,, \label{M} \\
  \Omega^\mu &\equiv & \gamma^\mu(c_V-c_A \gamma^5)\,,
  \\
  \Sigma^\mu&
  \equiv&
\gamma^\mu(1-\gamma^5)
\,.
 \end{eqnarray}
In (\ref{M}) we have used the condition $q^2 \ll M_W^2$, where $M_W$ is the mass of the
vector gauge boson $W$, with $q^\mu=p_n^\mu-p_p^\mu$  the transferred
momentum. From Eq. (\ref{rateproc1}) it follows that
 \begin{equation}\label{rateproc1fin}
    \lambda_{n+\nu_e\to p+e^-}=A \, {\cal T}^5 I_y\,,
 \end{equation}
where
 \begin{equation}
  A\equiv \frac{g_V+3g_A}{2\pi^3}\,,
 \end{equation}
and where
\begin{equation}
 I_y=\int_y^\infty \epsilon(\epsilon-{\cal Q}')^2\sqrt{\epsilon^2-y^2}\, n(\epsilon-{\cal
Q})[1-n(\epsilon)]d\epsilon,
 \end{equation}
 with
 \begin{equation}
 y\equiv \frac{m_e}{{\cal T}}\,, \quad {\cal Q}'=\frac{{\cal Q}}{{\cal T}}\,.
 \end{equation}

  A similar calculation for the process (\ref{proc2}) gives
  \begin{equation}\label{rateproc2fin}
    \lambda_{e^+ + n \to p+ {\bar \nu}_e}=A \, {\cal T}^5 J_y\,,
 \end{equation}
 with
 \begin{equation}
 J_y=\int_y^\infty \epsilon(\epsilon+{\cal Q}')^2\sqrt{\epsilon^2-y^2}\,
n(\epsilon)[1-n(\epsilon+{\cal Q}')]d\epsilon\,,
 \end{equation}
 which finally  results to
 \begin{equation}
 \label{ne-pnu-fin}
    \lambda_{e^+ + n\to p+{\bar \nu}_e}=A\, {\cal T}^3(4! {\cal T}^2+2\times 3! {\cal
Q}{\cal T}+2! {\cal
Q}^2)\,.
 \end{equation}

Lastly, for the neutron decay (\ref{proc3}) one obtains
 \begin{equation}\label{rateproc3}
    \tau=\lambda_{n\to p+e^- +{\bar \nu}_e}^{-1}\simeq 887 \text{sec}\,.
 \end{equation}
 Hence, in the calculation of (\ref{sumprocess}) we can safely neglect the above
interaction rate of the neutron decay, i.e. during the BBN the neutron can be
considered as a stable particle.

The above approximations (i)-(iii) lead to
\cite{bernstein}
 \begin{equation}
 \label{auxilirel}
  \lambda_{e^+ +n\to p+{\bar \nu}_e}=\lambda_{n+\nu_e\to p+e^-}\,.
 \end{equation}
 Thus, inserting   (\ref{auxilirel}) into (\ref{sumprocess}), and then
into (\ref{LambdaA}), allows to derive the expression for
$\Lambda({\cal T})$, namely
 \begin{equation}\label{LambdafinA}
    \Lambda({\cal T})\simeq 2\lambda_{np}=4\lambda_{e^+ +n\to p+{\bar \nu}_e}\,,
 \end{equation}
 which using (\ref{ne-pnu-fin}) leads to
  \begin{equation}\label{LambdafinApp}
    \Lambda({\cal T}) =4 A\, {\cal T}^3(4! {\cal T}^2+2\times 3! {\cal Q}{\cal T}+2!
{\cal Q}^2)\,.
 \end{equation}

\end{document}